\documentclass[preprintnumbers,article,amsmath,amssymb,floatfix,10pt,prd,onecolumn,
superscriptaddress,nofootinbib]{revtex4-2}
\usepackage{bm}
\usepackage{amsfonts}
\usepackage{latexsym}
\usepackage[latin1]{inputenc}
\usepackage{graphicx}
\usepackage{amsmath}
\usepackage{palatino}
\usepackage{mathpazo}
\usepackage{textcomp}
\usepackage{float}
\usepackage{booktabs}
\usepackage{dcolumn}
\usepackage{hyperref}
\hypersetup{colorlinks,citecolor=blue}
\hypersetup{colorlinks=true,linkcolor=red,filecolor=magenta, urlcolor=blue}
\usepackage{amsmath}
\usepackage{xcolor}
\usepackage{tabularx}
\usepackage{multirow}
\usepackage{mathtools}
\newcommand{\be}{\begin{equation}}
\newcommand{\ee}{\end{equation}}
\newcommand{\bea}{\begin{eqnarray}}
\newcommand{\eea}{\end{eqnarray}}
\newcommand{\eeas}{\end{eqnarray*}}
\newcommand{\beas}{\begin{eqnarray*}}
\begin{document}
\title{Models of $f(Q)$ gravity with electromagnetic field}
\author{S. H. Shekh}
\email{da\_salim@rediff.com}
\affiliation{Department of Mathematics. S. P. M. Science and Gilani Arts Commerce
College, Ghatanji, Dist. Yavatmal,\\ Maharashtra-445301, India.}
\author{Hira Sohail}
\email{hirasohail02@outlook.com}\affiliation{Centre for High Energy Physics, University of the Punjab, 54590, Lahore, Pakistan.}
\author{Irfan Mahmood}
\email{mahirfan@yahoo.com}\affiliation{Department of Mathematics, Shanghai University  and Newtouch Center for Mathematics of Shanghai University,  Shanghai ,200444, P.R.China.}\affiliation{Centre for High Energy Physics, University of the Punjab, 54590, Lahore, Pakistan.}
\author{Allah Ditta}
\email{mradshahid01@gmail.com}\affiliation{Department of Mathematics, Shanghai University  and Newtouch Center for Mathematics of Shanghai University,  Shanghai ,200444, P.R.China.}
\author{Anil Kumar Yadav}
\email{abanilyadav@yahoo.co.in}
\affiliation{Department of Physics, United College of Engineering and Research, Greater Noida -201310, India.}

\begin{abstract}
\textbf{Abstract:} There are so many ideas that potentially explain the dark energy phenomenon, current research is focusing on a more in-depth analysis of the potential effects of modified gravity on both local and cosmic scales. In this paper we have investigated some cosmic reconstructions in $f (Q)$ cosmology where $Q$ is the non-metricity corresponding to the evolution background in the Friedmann-Lamatre-Robertson-Walker $(FLRW)$ universe. This allows us to determine how any $FLRW$ cosmology can emerge from a particular $f (Q)$ theory. We employ the reconstruction technique to generate explicit formulations of the $f (Q)$ Lagrangian for several types of matter sources like perfect fluid, dust like fluid, stiff fluid and the binary mixture of two fluids. Furthermore, we computed the field equations and equation of state (EoS) parameter $\omega$ for two different reconstructed $f(Q)$ models with the variation of the involved constants, which gives the scenario of accelerating universe, quintessence region and cosmological constant. We also observed that the time dependence of  $\omega$ admits cosmic acceleration. These new $f(Q)$ gravity inspired models may have an impact on gravitational phenomena at other cosmological scales.
\end{abstract}
\maketitle
\date{\today}
%%%%%%%%%%%%%%%%%%
\section{Introduction}\label{intro}
One notable troubling issue during the previous two decades has been the universe presents acceleration due to the Dark Energy (DE). Recent advancements in observational cosmology confirm this cosmological occurrence; for example, type Ia supernovae \cite{A1,A2,A3}, Cosmic Microwave Background Radiation\cite{A4}, observations of large-scale structure \cite{A5,A6,A7}, the power spectrum of the Lyman-forest from the Sloan Digital Sky Survey\cite{B1}, and the exploration of high-energy DE models using weak lensing data \cite{B2}. On the other hand, GR cannot explain observations of huge pulsars \cite{B3,B4} and white dwarfs \cite{B5,B6,B7} with masses greater than the traditional maximum limit (i.e. the Chandrasekhar mass limit, $1.44M_{\circledcirc }$). Furthermore, the strong gravitational field and new findings contradict $GR$ \cite{C1,C2,C3}. As a result, scientists look for suitable changes to the General Theory of Relativity (GR) e.g, $f (R)$ gravity \cite{B8,B9,B10,B11,B12}, $f (R,G)$ gravity \cite{B13}, $f (T)$ gravity \cite{B14,B15,B16}, $f (G)$ gravity \cite{B17,B18,B19}, Brans-Dicke (BD) gravity \cite{B20,B21}, and so on, where $R, G,$ and $T$ are the Ricci, Gauss- Bonnet, and torsion scalars respectively. The theory of higher-order curvature, particularly $f (R)$ gravity, is the most successful modification of GR, explaining the presence of dark matter and using evidence to contradict the theory of gravity\cite{C4}. In the recent past, Jim'enez et al. \cite{D1} have suggested a symmetric teleparallel theory of gravity i. e. $f(Q)$ gravity, which is a well-motivated gravity theory in which non-metricity drives the gravitational interaction in space-time, based on Lagrangian density yielding a general function of the non-metricity scalar $Q$. At the background level, the modified theory of $f(Q)$ gravity leads to fascinating the cosmic phenomenology \cite{E1,E2,E3,E4,E5,E6,E7,E8,E9,E10,E11,E12,E13,E14,E15,E16,E17,E18,E19,E20}.\\ 

Furthermore, it has successfully been confronted with various background and perturbation observational data, such as the Cosmic Microwave Background (CMB), Supernovae type Ia (SNIa), Baryonic Acoustic Oscillations (BAO), Redshift Space Distortion (RSD), growth data, and so on \cite{E21,E22,E23,E24,E25,E26,E27}, revealing that $f(Q)$ gravity may challenge the standard $\Lambda CDM$ scenario. Finally, $f(Q)$ gravity easily overcomes the Big Bang Nucleosynthesis (BBN) limitations\cite{E28}.Choosing $T$ (torsion) or $Q$ (nonmetricity) produces similar but distinct theories of gravity, named $(TEGR)$ teleparallel equivalent of general relativity \cite{F1,F2} and (STGR) symmetric teleparallel general relativity\cite{F3,F4,F5}. The notion of gravity in STGR is based on nonmetricity rather than curvature and torsion.\\ 
Inspired by the interesting qualities of f $(Q)$ gravity, we hope to create a family of $f(Q)$ functions that can imitate the properties of the $\Lambda $CDM model.\\

In this research, we will investigate several specific $f(Q)$  gravity theories and rebuild the related cosmological theory for many cosmological solutions using a technique introduced in \cite{1,2}. Section \ref{intro} is introductory in nature. In section \ref{1}, we have write the basics of $f(Q)$ gravity and charged field equations for anisotropic matter  using the action in ,we also have analytically obtained the modified Friedmann equations using the charged field equations for FLRW metric. In the section \ref{2}, the cosmological models  is being developed starting form the $\Lambda CDM$ model for various kind of fluid.In the section \ref{3}, various cosmological models are developed for various values of EoS parameter $\omega$ using the modified Friedman equations.In the section \ref{4}, the values of energy density($\rho$) and pressures ($p_t \And p_r$) for known $f(Q)$ models are calculated.In the final section \ref{5} a cosmological reconstruction for modified f(Q) gravity is built in terms of e-folding and EoS parameter $\omega$.
%%%%%%%%%%%%%%%%%%%%%%%%%%%
\section{Basics of $f(Q)$ gravity and its field equations}\label{1}
Let us start with the following action of $f(Q)$ gravity
\begin{equation}\label{eq.1}
    S=\int \sqrt{-g} d^4x \Bigg[ \frac{1}{2}f(Q)+\lambda_{\alpha}^{\beta \mu \nu} R^{\alpha}_{\beta \mu \nu} +\lambda_{\alpha}^{\mu \nu}T_{\mu \nu}^{\alpha}+L_m+L_e\Bigg]
\end{equation}
where $f(Q)$ is the function of nonmetricity $Q$,$\hspace{0.2cm}\lambda_{\alpha}^{\beta \mu \nu}$ are the Lagrangian multiplier,$g$ indicates the determinant of metric element,$L_m$ and  $L_e$ denote the matter Lagrangian density and  electric field. In the context of affine-connection, non-metricity is defined as 
\begin{equation}
    Q_{\alpha \mu \nu}=\Delta_\alpha g_{\mu \nu}=\partial_\alpha  g_{\mu \nu}-\Gamma^{\varrho}_{\alpha \mu}g_{\varrho \nu}-\Gamma^{\varrho}_{\alpha \nu}g_{\mu \varrho}
\end{equation}
The commonly used form of the affine-connection can be divided into three components, which are listed below.

\begin{equation}
    \Gamma^{\varrho}_{\mu \nu}=\left\{\begin{array}{lr}
        \varrho& \\
        \mu \nu& \end{array}\right\}+K^{\varrho}_{\mu \nu}+L^{\varrho}_{\mu \nu}
\end{equation}
where the Levi-Civita $\left\{\begin{array}{lr}
        \varrho& \\
        \mu \nu& \end{array}\right\}$ can be defined in terms of metric $g_{\mu \nu}$ as
\begin{equation}
    \left\{\begin{array}{lr}
        \varrho& \\
        \mu \nu& \end{array}\right\}=\frac{1}{2}g^{\varrho \beta} \big( \partial_\mu g_{\beta \nu}+\partial_\nu g_{\beta \mu}-\partial_{\beta}g_{\mu \nu}\big) 
\end{equation}
where $K^{\varrho}_{\mu \nu}$ denotes the contortion specified by the following relation:
\begin{equation}
    K^{\varrho}_{\mu \nu}=\frac{1}{2}T^{\varrho}_{\mu \nu}+T_{\mu \hspace{0.2cm}\nu}^{\hspace{0.18cm}\varrho}
\end{equation}
Torsion tensor $T^{\varrho}_{\mu \nu}$ is realized as the antisymmetric component of the affine connection,$T^{\varrho}_{\mu \nu}=2\Gamma^{\varrho}_{[\mu \nu]}$ in the previous equation, and the disformation $ L^{\varrho}_{\mu \nu}$ is written as
\begin{equation}
    L^{\varrho}_{\mu \nu}=\frac{1}{2}Q^{\varrho}_{\mu \nu}-Q_{\mu \hspace{0.2cm}\nu}^{\hspace{0.18cm}\varrho}
\end{equation}
The non-metricity conjugate is calculated as
\begin{equation}
    P^{\alpha}_{\mu \nu}=-\frac{1}{4}Q^{\alpha}_{\mu \nu}+\frac{1}{2}Q^{\alpha}_{(\mu^{\alpha}\nu)}+\frac{1}{4}\Bigg(Q^\alpha -\Tilde{Q}^{\alpha}\Bigg)g_{\mu \nu}-\frac{1}{4}\delta^{\alpha} _{\hspace{0.2cm}(\mu}Q_{\nu )}
\end{equation}
with two independent traces
\begin{equation}
    Q_\alpha=Q_{\alpha \hspace{0.195cm}\mu}^{\hspace{0.18cm}\mu},\hspace{1cm}\Tilde{Q}_\alpha=Q^{\mu}_{\alpha \mu}
\end{equation}
The non-metricity scalar is calculated is as follows
\begin{equation}\label{eq.9}
    Q=-Q_{\alpha \mu \nu}P^{\alpha \mu \nu}
\end{equation}
The standard form of the energy-momentum tensor and electromagnetic field tensor is given by
\begin{equation}
    T_{\mu \nu}=-\frac{2}{\sqrt{-g}}\frac{\delta(\sqrt{-g}L_m)}{\delta g^{\mu \nu}}
\end{equation}
\begin{equation}
    E_{\mu \nu}=-\frac{2}{\sqrt{-g}}\frac{\delta(\sqrt{-g}L_e)}{\delta g^{\mu \nu}}
\end{equation}
 Varying the action in Eq.(\ref{eq.1}) with respect to the metric $g_{\mu \nu}$, we get the following field equations\cite{M}
\begin{equation}\label{eq.12}
    -T_{\mu \nu} +E_{\mu \nu}=\frac{2}{\sqrt{-g}}\Delta_\alpha \bigg(\sqrt{-g} f_Q P^{\alpha}_{\mu \nu}\bigg)+\frac{1}{2}g_{\mu \nu}f+F_Q \bigg(P_{\mu \alpha \beta} Q^{\hspace{0.15cm}\alpha \beta }_\nu -2 Q_{\alpha \beta \mu}P^{\alpha \beta}_{\hspace{0.4cm}\nu}\bigg)
\end{equation}

We assume a flat Friedmann-Lemaitre-Robertson-Walker
(FLRW) metric with $N(t)=1$
\begin{equation}\label{eq.13}
  ds^2=-N(t)^2 dt^2+a(t)^2 \sum_{i=1}^{3} (dx^i)^2
\end{equation}
where $N(t)$ is a Lapse function and $a(t)$ is a cosmic scale
factor.We cannot select a certain Lapse function because the coincident gauge is fixed during the diffeomorphism.
The particular case of $Q$ theories, on the other hand, allows for this since $Q$ retains a residual time re-parameterization invariant. As a result, one can use symmetry to set $N(t) = 1$ \cite{Gadbail2023}. The covariant derivatives are reduced to ordinary derivatives by using the coincident gauge.If we adopt a co$-$moving reference system,with $\zeta^\mu=\{0,0,0,1\}$,then $\frac{d}{d\tau}=\frac{d}{dt^{\prime}}$,$\tau$ is a proper time, giving $H=\frac{\Dot{a(t)}}{a(t)}$.\\
So, for the metric (\ref{eq.13}), the Eq.(\ref{eq.9})yields the non-metricity scalar Q as
\begin{equation}\label{14.1}
    Q=6H^2
\end{equation}
The energy-momentum tensor containing charge for
anisotropic matter source\cite{Bhar2019} is defined as
\begin{equation}\label{eq.15}
    T_{\mu \nu}=\rho \zeta_\mu \zeta_\nu +p_r \xi _\mu \xi_\nu+p_t (\zeta_\mu \zeta_\nu -\xi_\mu \xi_\nu-g_{ij})
\end{equation}
where,$ p_r$, and $p_t$ denote the energy density and pressure components (radial and tangential). The expression $\zeta_\mu$ defines the radial four-vector and $\xi_\mu$ elaborates the four-velocity vector, both of which meet the following condition,
\begin{equation}
    \zeta^\beta\zeta_\beta=1,\hspace{0.4cm} \xi^\beta \xi_\beta =-1
\end{equation}
In contrast, the  Electromagnetic tensor which is given by  Lichnerowicz \cite{AL1} is 
\begin{equation}\label{eq.19}
   E_{i}^j=4 \pi \Big[|h|^2\Big(\zeta_i \zeta^j+\frac{1}{2}g_j^i\Big)-h_ih^j\Big] 
\end{equation}
where $h_i$ is the magnetic flux defined by 
\begin{equation}
    h_i=\frac{\sqrt{-g}}{8 \pi} \epsilon_{ijkl}F^{kl}\zeta^j
\end{equation}
If we suppose that the current is flowing in the x-axis direction, the magnetic field generated will be in the $yz$-plane. As a consequence, $h_1 \neq 0,$ $h_2 = h_3 = h_4 = 0$.
Using Eq. (17), it is clear that $F_{23}$ will be the only  component of $F_{ij}$ which is non-vanishing and all other components$ (F_{12}, F_{13}, F_{14}, F_{24},F_{34})$ vanish save $F_{23}$, which is non-vanishing. Because the magnetic field vector indicates a preferred spatial direction, any cosmological model that includes a global magnetic field must be anisotropic. As a result, the Maxwell equations
\begin{equation} F_{ij,k}+F_{jk,i}+F_{ki,j}=0
\end{equation}
gives
\begin{equation}
    F_{23}=constant=M
\end{equation}
We get the components of the stress tensor and electromagnetic tensor for the line element (\ref{eq.13})
as
\begin{equation}\label{21.1}
    T_{11}=a(t)^2 p_r,\hspace{0.4cm}T_{22}=T_{33}=a(t)^2p_t,\hspace{0.4cm}T_{44}=\rho
\end{equation}
\begin{equation}\label{22.1}
    E_{11}=-\frac{M^2}{8 \pi a^2},\hspace{0.4cm}E_{22}=E_{33}=\frac{M^2}{8 \pi a^2},\hspace{0.4cm}E_{44}=\frac{M^2}{8 \pi a^4}
\end{equation}
We find the field equations given below by using the equations (\ref{21.1}), (\ref{22.1}) and (\ref{14.1}) as

\begin{equation}
    \rho=\frac{1}{2}f-6 H^2 f_Q-\mathbb{E}^2
\end{equation}
\begin{equation}
    p_t=2 \Dot{f_Q}H+6f_Q H^2+2f_Q \Dot{H}-\frac{1}{2}f-\mathbb{E}^2
\end{equation}
\begin{equation}
    p_r=2 \Dot{f_Q}H+6f_Q H^2+2f_Q \Dot{H}-\frac{1}{2}f+\mathbb{E}^2
\end{equation}
where $\mathbb{E}^2=\frac{-M^2}{8 \pi a^4}$.\\

The equation of continuity reads as
\begin{equation}
    \Dot{\rho}+3H(\rho+p_r)=0
\end{equation}
Thus the modified Friedmann equations can be rewritten as\begin{equation}\label{eq.24}
    \rho=\frac{1}{2}f-Qf_Q-\mathbb{E}^2
\end{equation}
\begin{equation}\label{eq.25}
    p_t=2 \Dot{f_Q}H+f_QQ+2f_Q \Dot{H}-\frac{1}{2}f-\mathbb{E}^2
\end{equation}
\begin{equation}\label{eq.26}
    p_r=2 \Dot{f_Q}H+f_QQ+2f_Q \Dot{H}-\frac{1}{2}f+\mathbb{E}^2
\end{equation}

The parameter of state in this case is given by
\begin{equation}\label{eq.27}
    \omega=\frac{p_r}{\rho}
\end{equation}
By using Eq.(\ref{eq.24}) and (\ref{eq.26}) the state parameter can also be written as
\begin{eqnarray}\label{eq.28}  \omega=-1+\frac{2\Dot{H}(12\Dot{f_QQ}+f_Q )}{-(Qf_Q-\frac{1}{2}f+\mathbb{E}^2)}  
\end{eqnarray} In the de Sitter universe, where $\Dot{H} = 0 $, Eq. (\ref{eq.28}) produces $\omega =- 1$, which is consistent with the $\Lambda CDM$ model. So, we can construct the $f(Q)$ theory for $\Lambda CDM $ model.

\section{Reconstruction of $f(Q)$ theory: $\Lambda$CDM model }\label{2}
Starting with the $\Lambda$CDM model, let us reconstruct the $f(Q)$ theories, as is well known, in this instance
\begin{equation}\label{eq.29}
    H^2=H_0^2+\frac{\kappa^2 \rho_0}{3 a^3}=\frac{1}{6}Q
\end{equation}
with $\kappa^2=1 $,as assumed in this case.The equation (\ref{eq.29}) can also be written as
\begin{equation}
   Q=6H_0^2+\frac{2 \rho_0}{a^3},\hspace{0.5cm} a^{-1}=\Bigg[\frac{3}{ \rho_0}(H^2-H_0^2)\Bigg]^{\frac{1}{3}}=\Bigg[\frac{1}{2 \rho_0}(Q-6H_0^2)\Bigg]^{\frac{1}{3}}
\end{equation}
The deceleration parameter $(q)$ is defined as
\begin{equation}\label{q.1}
 q=-\frac{a \Ddot{a}}{\Dot{a}^2}   
\end{equation}
The value of $q$ for this model is calculated below, where $c_1= constant$
\begin{equation}\label{q-1}
    q=-1+\frac{3}{1+\cosh \left(3 H_0 \left(t-\sqrt{3} c_1 \right)\right)}
\end{equation}
The solution to equation (\ref{eq.24})'s homogeneous case (that is, $\rho= 0$) is given by
\begin{equation}\label{34.1}
    f=2 \mathbb{E}^2+C \sqrt{Q}, \hspace{0.5cm} C=const.
\end{equation}
The second equation (\ref{eq.26}) naturally satisfies and yields  $p_r=0$, and the equation (\ref{eq.25}) for $f(Q)$  yields $p_t=-2 \mathbb{E}^2$. Thus, giving the null dust solutions ($\rho=0, p_r=0$).The phenomenons which can be modeled by the solutions include  a very high-frequency electromagnetic wave, and a beam of incoherent electromagnetic radiation. The model in (\ref{34.1}) is a real-valued function with a positive non-metricity scalar, as shown by the observation. It means that, in addition to GR, there are classes of real-valued functions $f (Q)$ that can simulate a dust-like expansion history.\\

The equation of state parameter in terms of energy density can also be derived form the equation of continuity as
\begin{equation}\label{eq.32}
    \rho=\rho_c a^{-3(1+\omega)}
\end{equation}
Let us now construct the $f(Q)$ gravity theories for matter field 
\subsection{Reconstruction for dust-like matter}
First, consider the situation in which the universe is filled with dust-like things for which the value of equation of state is $\omega = 0$. By imposing the dust like case, we obtained the energy density from the equation (\ref{eq.32}) in terms of $a$ as

\begin{equation}
    \rho=\rho_c a^{-3}=\frac{\rho_c}{2\rho_0}(Q-6H^2_0)
\end{equation}
Substituting the above equation in (\ref{eq.24}), we get the following equation with unknown function $f(Q)$
\begin{equation}
    \frac{1}{2}f(Q)-Q f_Q-\mathbb{E}^2=\frac{\rho_c}{2 \rho_0}\Bigg(Q-6H_0^2\Bigg)
\end{equation}
The particular solution of the above equation is 
\begin{equation}
    f(Q)=2\mathbb{E}^2-\frac{\rho_c}{\rho_0}\Bigg(Q+6 H_0^2\Bigg)
\end{equation}
The value of deceleration parameter for this can be calculated using Eq. (\ref{q.1}) as
\begin{equation}
    q = -1
\end{equation}
\subsection{Reconstruction for Dark Matter  fluid}
Now we consider the perfect fluid with the equation of state parameter value as $\omega=-\frac{1}{3}$. The cosmos is accelerating in this scenario, and the EoS value $\omega=-\frac{1}{3}$ is physically interesting since it is close to the limit of the set of matter fields that satisfy the strong energy condition.Also,it is shown in \cite{Vinutha2023} that the negative strong energy condition shows cosmic acceleration due to anti gravity. From the equation (\ref{eq.32}), we found the energy density in terms of $a$ as,
\begin{equation}
    \rho=\rho_c a^{-2}=\rho_c\Bigg[\frac{1}{2 \rho_0}(Q-6 H_0^2)\Bigg]^{\frac{2}{3}}
\end{equation}
The modified Friedmann equation (\ref{eq.24}) takes the form 
\begin{equation}
    \frac{1}{2}f(Q)-Q f_Q-\mathbb{E}^2=\rho_c \Bigg[\frac{1}{2 \rho_0}\Bigg(Q-6H_0^2\Bigg)\Bigg]^{\frac{2}{3}}=\alpha \Bigg(Q-\beta\Bigg)^{\frac{2}{3}}
\end{equation}
The solution of the above equation is given by
\begin{equation}\label{41.1}
    f(Q)=2 \Bigg[\mathbb{E}^2+\frac{\alpha(Q-\beta)^{2/3}Hypergeomertic2F1\big[-\frac{2}{3},-\frac{1}{2},\frac{1}{2},\frac{Q}{\beta}\big]}{\big(1-\frac{Q}{\beta}\big)^{2/3}}\Bigg]
\end{equation}
The obtained explicit form of the $f (Q)$ ,in the  model(\ref{41.1}) is useful for investigating the universe's acceleration scenario, along with the positive non-metricty $Q$.
The value of Deceleration parameter in this case is 
\begin{equation}
    q=-1
\end{equation}
\subsection{Reconstruction for  stiff fluid}
Now we consider the stiff fluid with the equation of state parameter value as $\omega = 1$. The string theory predicts that a stiff cosmic fluid with pressure equal to the energy density can be characterised by a mass-less scalar field. The stiff fluid, on the other hand, is a crucial element since it dominates the remaining parts of the model early on and can be used to explain the shear-dominated phase of a potential initial anomalous scenario\cite{AT1} . Also,  A stiff matter era is present in the cosmological model of Zeldovich (1972) where the primordial universe is assumed to be made of a cold gas of baryons\cite{AT2}. Thus, equation (\ref{eq.32}) takes the form
\begin{equation}
    \rho=\rho_s a^{-6}=\rho_s\Bigg[\frac{1}{2 \rho_0}\Bigg(Q-6H_0^2\Bigg)\Bigg]^2,\hspace{0.5cm} \rho_s=const.
\end{equation}
In this case the modified equation(\ref{eq.24}) becomes
\begin{equation}
    \frac{1}{2}f(Q)-Q f_Q-\mathbb{E}^2=\rho_s \Bigg[\frac{\rho_0}{2 \rho_0}\Bigg(Q-6H_0^2\Bigg)\Bigg]^2=\alpha \Bigg(Q-\beta\Bigg)^2
\end{equation}
with the solution as
\begin{equation}
    f(Q)=2\mathbb{E}^2+\frac{2\alpha}{3}\Bigg(6Q\beta-Q^2+3 \beta^2\Bigg)
\end{equation}
\subsection{Reconstruction for two fluids}
The interactions of various "matter" components, which largely interact through gravity and electromagnetic radiation, are the foundation of our current understanding of the Universe. The idea of coupled ideal fluids is frequently used to describe the nature of the various components and potential interactions. We begin with two fluid species as opposed to one. With current densities of $\rho_c$ and $\rho_s$, respectively, let's assume that the cosmos contains both perfect fluid and stiff fluid.\\
The equation (\ref{eq.32}) is used in this instance to obtain the total matter density as 
\begin{equation}
\rho=\rho_c a^{-2}+\rho_s a^{-6}
\end{equation}
\subsubsection{Dark Matter fluid and stiff fluid}
Now let us consider the more complicated case with two fluids , perfect ($\omega=\frac{-1}{3}$) and stiff($\omega=1$), for which the equation (\ref{eq.32}) becomes 
\begin{equation}
    \rho=\rho_c a^{-2}+\rho_s a^{-6}=\rho_c\Bigg[\frac{1}{2 \rho_0}(Q-6 H_0^2)\Bigg]^{\frac{2}{3}}+\rho_s\Bigg[\frac{1}{2 \rho_0}\Bigg(Q-6H_0^2\Bigg)\Bigg]^2
\end{equation}
the equation (\ref{eq.24}) takes the form
\begin{equation}
    \frac{1}{2}f(Q)-Q f_Q-\mathbb{E}^2=\rho_c\Bigg[\frac{1}{2 \rho_0}(Q-6 H_0^2)\Bigg]^{\frac{2}{3}}+\rho_s \Bigg[\frac{1}{2 \rho_0}\Bigg(Q-6H_0^2\Bigg)\Bigg]^2=\alpha \Bigg(Q-\beta\Bigg)^{\frac{2}{3}}+\delta\Bigg(Q-\beta\Bigg)^2
\end{equation}
which has the solution
\begin{equation}
    f(Q)=\frac{2}{3}\Bigg[3\mathbb{E}^2-\alpha Q^2+6\alpha \beta Q+3\alpha \beta^2+\frac{3 \delta(Q-\beta)^{2/3}Hypergeomertic2F1\big[-\frac{2}{3},-\frac{1}{2},\frac{1}{2},\frac{Q}{\beta}\big]}{\big(1-\frac{Q}{\beta}\big)^{2/3}}\Bigg]
\end{equation}
where $\alpha=\frac{\rho_c}{2\rho_0},\hspace{0.2cm}\beta=6H^2_0,\hspace{0.2cm}\delta=\frac{\rho_s}{2\rho_0}$\\
\vspace{1cm}
\subsubsection{Dust like fluid and stiff fluid}
Now we consider the following two fluid model: dust like fluid ($\omega=0$) and stiff fluid($\omega=1$), which gives the following form for equation(\ref{eq.32})
\begin{equation}
    \rho=\rho_c a^{-3}+\rho_s a^{-6}=\rho_c\Bigg[\frac{1}{2 \rho_0}(Q-6 H_0^2)\Bigg]+\rho_s\Bigg[\frac{1}{2 \rho_0}\Bigg(Q-6H_0^2\Bigg)\Bigg]^2=\alpha \Bigg(Q-\beta\Bigg)+\delta\Bigg(Q-\beta\Bigg)^2 
\end{equation}
the equation (\ref{eq.24}) takes the form
\begin{equation}
     \frac{1}{2}f(Q)-Q f_Q-\mathbb{E}^2=\alpha \Bigg(Q-\beta\Bigg)+\delta\Bigg(Q-\beta\Bigg)^2 
\end{equation}
with the solution as
\begin{equation}
    f(Q)=\frac{2}{3}\Bigg[3\mathbb{E}^2-\alpha Q^2+6\alpha \beta Q+3\alpha \beta^2-3 \delta Q-3\delta \beta\Bigg]
\end{equation}
\vspace{1cm}
\subsubsection{Radiation  fluid and stiff fluid}
At last, we consider the following two fluid model:dust like fluid ($\omega=0$) and Radiation fluid($\omega=\frac{1}{3}$), which gives the following form for equation(\ref{eq.32})

\begin{equation}
    \rho=\rho_c a^{-3}+\rho_s a^{-4}=\rho_c\Bigg[\frac{1}{2 \rho_0}(Q-6 H_0^2)\Bigg]+\rho_s\Bigg[\frac{1}{2 \rho_0}\Bigg(Q-6H_0^2\Bigg)\Bigg]^{\frac{4}{3}}=\alpha \Bigg(Q-\beta\Bigg)+\delta\Bigg(Q-\beta\Bigg)^{\frac{4}{3}} 
\end{equation}
the modified Friedmann equation (\ref{eq.24}) takes the form
\begin{equation}
    \frac{1}{2}f(Q)-Q f_Q-\mathbb{E}^2=\rho_c\Bigg[\frac{1}{2 \rho_0}(Q-6 H_0^2)\Bigg]+\rho_s\Bigg[\frac{1}{2 \rho_0}\Bigg(Q-6H_0^2\Bigg)\Bigg]^{\frac{4}{3}}=\alpha \Bigg(Q-\beta\Bigg)+\delta\Bigg(Q-\beta\Bigg)^{\frac{4}{3}}
\end{equation}
with solution
\begin{equation}
    f(Q)=-\frac{2}{Z}\Bigg[Z\big(-\mathbb{E}^2+\delta(Q+\beta)\big)-\alpha \beta^2Z Y-\alpha \beta Q Z W\Bigg]
\end{equation}
where 
\begin{eqnarray*}
    Y&=&Hypergeomertic2F1\big[-\frac{1}{2},-\frac{1}{3},\frac{1}{2},\frac{Q}{\beta}\big] \hspace{2cm}
 Z=\Bigg(1-\frac{Q}{\beta}\Bigg)^{\frac{1}{3}}\\
  W&=&Hypergeomertic2F1\big[-\frac{1}{3},\frac{1}{2},\frac{3}{2},\frac{Q}{\beta}\big]
\end{eqnarray*}
\section{Reconstruction using $\omega$}\label{3}
Though different theoretical approaches are there to explain the phenomenon of cosmic acceleration, till now none of them is definitely known as the appropriate one. The present trend of modelling of late-time cosmic acceleration is called reconstruction, where the model is built up by taking the observational data directly into account. This is actually the reverse way of finding the suitable cosmological model.By using the same concept  we will reconstruct $f(Q)$ for the particular values of the state parameter $\omega$, which can be the possible candidate for accelerating universe with the positive non-metricty $Q$.The following equation can easily be obtained from equation (\ref{eq.28}) as
\begin{equation}\label{eq.51}
    2 \Dot{f_Q}H+\big[(\omega+1)Q+2\Dot{H}\big]f_Q+(\omega+1)\big(\mathbb{E}^2-\frac{1}{2}f\big)=0
\end{equation}
\subsection{With $\omega=-1$ for $\Lambda$CDM model}
In this case equation (\ref{eq.51}) becomes
\begin{equation}
    \Dot{H}\Bigg[12 f_{QQ}H^2+f_Q\Bigg]=0
\end{equation}
There are two solutions to this equation. i) $\Dot{H} = 0$. As a result, $H = H_0 = const.$ It's called the de Sitter spacetime. ii) The second solution is consistent with the following equation
\begin{equation}
    12 f_{QQ}H^2+f_Q=0
\end{equation}
which has the solution
\begin{equation}
    f(Q)=-12 e^{-1/2}2QC_1+C_2,\hspace{0.6cm}C_j=consts.
\end{equation}
\subsection{$\omega=-1/3$}
In this case the equation (\ref{eq.51}) becomes
\begin{equation}
    2 f_{QQ}H^2+\Bigg[\frac{2}{3}Q+2\Dot{H}\Bigg]f_Q+\frac{2}{3}\Bigg[\mathbb{E}^2-\frac{1}{2}f\Bigg]=0
\end{equation}
which has the solution
\begin{equation}
    f(Q)=2\mathbb{E}^2+C_1e^{-Y}HermiteH\Bigg[-\frac{3}{2},Z\Bigg]+C_2 e^{-Y}Hypergeometric1F1\Bigg[\frac{3}{4},\frac{1}{2},Z^2\Bigg]
\end{equation}
where
\begin{equation}
    Y=\frac{Q^2+6Q \Dot{H}}{Q},\hspace{1cm}Z=\frac{Q}{\sqrt{6}H}+\frac{\sqrt{\frac{3}{2}}\Dot{H}}{H}
\end{equation}
\subsection{$\omega=0$}
For this case pressure is zero and equation (\ref{eq.51}) takes the form with solution as given below
\begin{equation}
    2 f_{QQ}H^2+\Bigg[Q+2\Dot{H}\Bigg]f_Q+\mathbb{E}^2-\frac{1}{2}f=0
\end{equation}
\begin{equation}
    f(Q)=2\mathbb{E}^2+C_1 e^{-Y}HermiteH\Bigg[-\frac{3}{2},Z\Bigg]+C_2e^{-Y}Hypergeometric1F1\Bigg[\frac{3}{4},\frac{1}{2},Z^2\Bigg] 
\end{equation}
where
\begin{equation}
    Y=\frac{3Q^2+12Q\Dot{H}}{2Q},\hspace{1cm} Z=\frac{Q}{2H}+\frac{\Dot{H}}{H}
\end{equation}
\section{$f(Q) models$}\label{4}
Now, we consider the following assumed $f(Q)$ models and find the resulting expressions for $\rho$ and $p_r$
\subsection{$f(Q)=Q-\alpha \gamma\big(1-\frac{1}{\exp{\frac{Q}{\gamma}}}\big)$}
In the first case, we consider the following model\cite{N1}
\begin{equation}
    f(Q)=Q-\alpha \gamma\big(1-\frac{1}{\exp{\frac{Q}{\gamma}}}\big)
\end{equation}
Where $\alpha \in (0,1)$ and $\gamma = \Omega H^2_0$, $\Omega$ is a dimensionless parameter and $H^2_0$ is  Hubble parameter

The modified Friedmann equations(\ref{eq.24}) and (\ref{eq.26})  with the above model becomes
\begin{equation}
    \rho=-\frac{Q}{2}-\frac{\alpha \gamma}{2}+\alpha Q e^{- Q/ \gamma} +\frac{\alpha \gamma}{2} e^{- Q/ \gamma}-E^2
\end{equation}
\begin{equation}
    p_r=\frac{Q}{2}+\frac{\alpha \gamma}{2}+2 \Dot{H}+\frac{4 \alpha}{\gamma}Q\Dot{H}e^{- Q/ \gamma}-\alpha Q e^{- Q/ \gamma}-2 \alpha \Dot{H}e^{- Q/ \gamma}-\frac{\alpha \gamma}{2}e^{- Q/ \gamma}+E^2
\end{equation}
The EoS paramter $\omega$ form equation (\ref{eq.27}) in this case takes the form
\begin{equation}
    \omega=-1-\frac{\frac{4 \alpha}{\gamma}Q\Dot{H}e^{- Q/ \gamma}-2 \alpha \Dot{H}e^{- Q/ \gamma}+2\Dot{H}}{\frac{Q}{2}+\frac{\alpha \gamma}{2}-\alpha Q e^{- Q/ \gamma} -\frac{\alpha \gamma}{2} e^{- Q/ \gamma}+E^2}=-1-\frac{A}{B}
\end{equation}
This results in three separate examples of $\omega$ representing different stages of the universe's evolution, as follows:
\begin{itemize}
    \item if $\frac{A}{B} <0$, then $\omega <-1$ which corresponds to the phantom accelerating universe
    \item if $0< \frac{A}{B}<1$,then EoS parameter $\omega$ will be slightly greater than $- 1$ which means that the universe stays in the quintessence region.
     \item if $\frac{A}{B}=0$,we have a cosmos whose dynamics is determined by the cosmological constant $\omega=-1$.

\end{itemize}

\subsection{$f(Q)=Q+k Q^n$}
For this case we consider the below model\cite{N2} with $k$  and $n$ as constants
\begin{equation}
    f(Q)=Q+kQ^n
\end{equation}
For this model the corresponding equations become
\begin{equation}\label{eq.66}
    \rho=-\frac{Q}{2}+\frac{k Q^n}{2}-nk Q^n-E^2
\end{equation}
\begin{equation}\label{eq.67}
    p_t=\frac{Q}{2}+nkQ^n-\frac{kQ^n}{2}+2\Dot{H}+2n(2n-1)k\Dot{H}Q^{n-1}-E^2
\end{equation}
\begin{equation}\label{eq.68}
    p_r=\frac{Q}{2}+nkQ^n-\frac{kQ^n}{2}+2\Dot{H}+2n(2n-1)k\Dot{H}Q^{n-1}+E^2
\end{equation}
The EoS paramter for the current modle is
\begin{equation}\label{eq.69}
    \omega=-1-\frac{2\Dot{H}+2n(2n-1)k \Dot{H}Q^{n-1}}{\frac{Q}{2}-\frac{kQ^n}{2}+nkQ^n+E^2}=-1-\frac{A}{B}
\end{equation}
For any positive real number $n \And k$, we can discuss as follows
\begin{itemize}
    \item if $\frac{A}{B} <0$, then $\omega <-1$ which corresponds to the phantom accelerating universe
    \item if $0< \frac{A}{B}<1$,then EoS parameter $\omega$ will be slightly greater than $- 1$ which means that the universe stays in the quintessence region.
     \item if $\frac{A}{B}=0$,we have a cosmos whose dynamics is determined by the cosmological constant $\omega=-1$.
\end{itemize}
Now, we study the current model for different values of constants $k$ and $n$
\subsubsection{k=0}
The equations (\ref{eq.66},\ref{eq.67},\ref{eq.68})and (\ref{eq.69}) with $k=0$ becomes
\begin{equation}
    \rho=-\frac{Q}{2}-E^2
\end{equation}
\begin{equation}
    p_t=\frac{Q}{2}+2\Dot{H}-E^2
\end{equation}
\begin{equation}
    p_r=\frac{Q}{2}+2\Dot{H}+E^2
\end{equation}
\begin{equation}
    \omega=-1+\frac{2 \Dot{H}}{-(\frac{Q}{2}+E^2)}
\end{equation}
\subsubsection{$k \neq 0$}
 In this section the Eqs. (\ref{eq.66}), (\ref{eq.67}), (\ref{eq.68}) and (\ref{eq.69}) for different value of $n$ with $k \neq 0$ is being discussed
i)For $n=1/2$, the Eqs. (\ref{eq.66}), (\ref{eq.67}), (\ref{eq.68}) and (\ref{eq.69}) becomes
\begin{equation}
    \rho=-\frac{Q}{2}-E^2
\end{equation}
\begin{equation}
    p_t=\frac{Q}{2}+2\Dot{H}-E^2
\end{equation}
\begin{equation}
    p_r=\frac{Q}{2}+2\Dot{H}+E^2
\end{equation}
\begin{equation}
    \omega=-1+\frac{2 \Dot{H}}{-(\frac{Q}{2})+E^2}
\end{equation}
ii)For  $n=2$, then the equations(\ref{eq.66},\ref{eq.67},\ref{eq.68},\ref{eq.69}) takes the form
\begin{equation}
    \rho=-\frac{Q}{2}-\frac{3}{2}kQ^2-E^2
\end{equation}
\begin{equation}
    p_t=\frac{Q}{2}+\frac{3}{2}kQ^2+2\Dot{H}+12k \Dot{H}Q-E^2
\end{equation}
\begin{equation}
    p_r=\frac{Q}{2}+\frac{3}{2}kQ^2+2\Dot{H}+12k \Dot{H}Q+E^2
\end{equation}
\begin{equation}
    \omega=-1+\frac{2\Dot{H}+12k\Dot{H}Q}{-(\frac{Q}{2}+\frac{3}{2}kQ^2+E^2)}
\end{equation}
\section{Cosmic acceleration}\label{5}
In this section, we discuss the evolution of scale factor in terms of EoS parameter for which we consider the following model\cite{2}
\begin{equation}\label{eq.82}
    p_r=\frac{A_{-1}(Q)}{\rho}+A_0(Q)+A_1(Q)\rho     
\end{equation}
It is worthwhile to  note that for $A_{-1} = constant, A_0 = A_1 = 0,$ Eq. (\ref{eq.82}) converts to the Chaplygin gas equation. Moreover, we obtain Eq. (\ref{eq.82}), for $f(Q)=Q+k Q^n$. Therefore, the expression for energy density is read as
\begin{equation}
    \rho=-\frac{Q}{2}+\frac{k Q^n}{2}-nk Q^n-\mathbb{E}^2
\end{equation}
\begin{equation}
    p_r=\frac{Q}{2}+nkQ^n-\frac{kQ^n}{2}+2\Dot{H}+2n(2n-1)k\Dot{H}Q^{n-1}+\mathbb{E}^2
\end{equation}
with the assumption that $n=2,k=1 \And A_k=cosnts.$\\ 
Eq. (\ref{eq.82}) takes the form
\begin{equation}
\Dot{H}=-\frac{3 \left(A_1+1\right) Q^2-2 A_0+\frac{4 A_{-1}}{2 \mathbb{E}^2+3 Q^2+Q}+2 \mathbb{E}^2+Q}{4 (6 Q+1)}
\end{equation}
or
\begin{equation}
    Q_N=-\frac{3 \left(3 \left(A_1+1\right) Q^2-2 A_0+\frac{4 A_{-1}}{2 \mathbb{E}^2+3 Q^2+Q}+2 \mathbb{E}^2+Q\right)}{6 Q+1}
\end{equation}
where $N=\log{\frac{a}{a_0}}$\\
As a result, for the scenario $A_0 = A_1 = 0$, we get the following solution, where $b=4 A_{-1}$
\begin{equation}
    Q_N=-\frac{3 \left(\frac{4 A_{-1}}{2 \mathbb{E}^2+3 Q^2+Q}+2 \mathbb{E}^2+3 Q^2+Q\right)}{6 Q+1}=-\frac{3 \left(\frac{b}{2 \mathbb{E}^2+3 Q^2+Q}+2 \mathbb{E}^2+3 Q^2+Q\right)}{6 Q+1}
\end{equation}
\begin{equation}\label{eq.88}
    Q=\frac{1}{6} \left(-\sqrt{-12 \sqrt{e^{-6(N-N_0)}-b}-24 \mathbb{E}^2+1}-1\right)
\end{equation}
or
\begin{equation}\label{eq.89}
    H=\frac{1}{6} \sqrt{-\sqrt{-12 \sqrt{e^{-6(N-N_0) }-b}-24 \mathbb{E}^2+1}-1}
\end{equation}
For the models (\ref{eq.88}-\ref{eq.89}), the EoS parameter($\omega$) takes the form
\begin{equation}\label{eq.90}
    \omega=-\frac{4 (6 Q+1) H'(t)}{2 \mathbb{E}^2+3 Q^2+Q}-1=-\frac{(6 Q+1) Q_N}{3 \left(2\mathbb{E}^2+3 Q^2+Q\right)}-1=-1+\frac{1}{1-b e^{6(N-N_0)}}
\end{equation} 
It is evident from Eq. (\ref{eq.90}) that the EOS parameter evolves in the range $-1 \leq \omega \leq 0 $ or $\omega \leq -1$. So, we can conclude that the some models presented in this paper, admit cosmic acceleration.
\section{Conclusion}
In this paper, we have investigated several forms of DE cosmology in $f(Q)$ gravity and examined its nature with electromagnetic field. We considered the well know relation between nonmetricty $Q$ and Hubble parameter in the form $Q = 6H^{2}$ and reconstructed $f(Q)$ theory for dust like matter, perfect fluid, stiff fluid and binary mixture of two fluids. Furthermore we also reconstructed $f(Q)$ theory by using various form of EOS parameter $\omega$. It is worthwhile to note that $\omega = -1$ generates a physically viable form of $f(Q)$ that describes the acceleration of the universe at present epoch. The various modified form of $f(Q)$ models are investigated which give a extremely powerful theory without need of DE components to describes the late time acceleration of the universe. Furthermore, we observe that the reconstruction of $f(Q)$ theories define the deceleration parameter $q$ as a function of $t$ and its explicit expression is obtained. From Eq. (\ref{q-1}), it is clear that $q$ varies with time $t$ and it approaches to -1 for dust like matter and Dark matter fluid; incidentally this value of $q$ leads to $\frac{dH}{dt} = 0$, which implies the greatest value of Hubble's parameter and the fastest rate of expansion of the universe. Therefore, the derived models under the reconstruction of $f(Q)$ theories can be utilized to describe the late time evolution of the actual universe. Finally, we conclude that no DE is required to replicate ordinary $\Lambda CDM$ cosmology and thus models of $f(Q)$ gravity with electromagnetic field are free from the problems associated with cosmological constant/DE as discussed in Refs. \cite{Abdalla/2022,Valentino/2021a,Valentino/2021b,Valentino/2021c,Valentino/2021d}. 
 %%%%%%%%%%%%%%%%%%%%%%%%%%%%%%%%%%%%%%%%%%%%%%%%%%%

\end{document}